\documentclass[12pt,a4paper]{article}
\usepackage{amsfonts}
\usepackage{amsmath}
\usepackage{amssymb}
\usepackage{latexsym}
\usepackage{color}
\usepackage{subfig}
\usepackage{relsize}
\usepackage{exscale} 
\usepackage[dvips]{graphicx}

\numberwithin{equation}{section}

\newcommand{\BbbR}{\mathbb{R}}
\newcommand{\BbbZ}{\mathbb{Z}}

\newcommand{\lscale}{{\bar{\mu}}}

\title{Singularity resolution by lattice shifts\\ 
in discretised quantum mechanics}

\author{Jorma Louko\thanks{jorma.louko@nottingham.ac.uk}, 
\ 
Samuel P. Philpott\thanks{ppysp1@nottingham.ac.uk} 
\ 
and 
Matthew D. Waller\thanks{pmymdw@nottingham.ac.uk}
\\
\noalign{\vspace{3ex}}
\small{\it School of Mathematical Sciences,
University of Nottingham,}\\
\small{\it Nottingham NG7 2RD, UK}
\\[3ex]
\small{(November 2013)}
\\[3ex]
\small{Published in Phys.\ Rev.\ D {\bf 89}, 044032 (2014)}
}

\date{}

\begin{document}

\maketitle

\begin{abstract}
We investigate the robustness of singularity avoidance mechanisms 
in nonrelativistic quantum mechanics on the discretised real line when 
lattice points are allowed to approach a singularity 
of the classical potential. 
We consider the attractive Coulomb potential and the 
attractive scale invariant potential, 
on an equispaced parity-noninvariant lattice 
and on a non-equispaced parity-invariant lattice, and we 
examine the energy eigenvalues by a combination 
of analytic and numerical techniques. 
While the lowest one or two eigenvalues descend to 
negative infinity in the singular limit, 
we find that the higher eigenvalues remain finite and form degenerate pairs, 
close to the eigenvalues of a 
theory in which a lattice point at the singularity 
is regularised either by Thiemann's 
loop quantum gravity singularity avoidance prescription or by a 
restriction to the odd parity sector. 
The approach to degeneracy can be 
reproduced from a nonsingular discretised 
half-line quantum theory by tuning a boundary condition parameter. 
The results show that Thiemann's singularity avoidance prescription 
and the discretised half-line boundary condition 
reproduce quantitatively correct features 
of the singular limit spectrum 
apart from the lowest few eigenvalues. 
\end{abstract}

\newpage


\newpage 

\section{Introduction}
\label{sec:intro}

In nonrelativistic quantum mechanics, a system whose classical
evolution is incomplete due to singularities in the potential may
yield a quantum theory with unitary time evolution provided the
singularities are sufficiently
weak~\cite{reed-simonII,blabk,extensions}. A~celebrated example is
the attractive Coulomb potential on~$\BbbR^3$, proportional
to~$-1/|{\bf x}|$, whose quantum theory is unitary if a boundary
condition at the origin is specified for the spherically symmetric
sector~\cite{fewster}.  Another example is the attractive scale
invariant potential on~$\BbbR_+$, proportional to~$-1/x^2$, where the
quantum theory is unitary if a boundary condition is specified
at the origin~\cite{narnhofer}.

In this paper we address systems with a singular potential within 
a quantisation framework 
that introduces a discrete lattice on the
classical configuration space. 
We shall examine the spectrum in the limit in which  
one or more lattice points approach the singularity. 

While this question may be of pragmatic interest when the discrete lattice 
is adopted as a practical tool for approximating 
a standard Schr\"odinger quantisation, 
our main motivation comes from the context in which the discreteness 
is fundamental, and specifically from 
the polymer quantisation framework \cite{afw,halvorson} that has been
employed in loop quantum gravity~\cite{Thiemann07,Rovelli07}. 
If a lattice point is taken to coincide with the classical 
singularity, the discrete quantum theory may be 
defined by explicitly regularising the 
potential~\cite{Thiemann,coulomb,scale-invariant}, 
or by restriction to the odd parity sector~\cite{scale-invariant}, 
or by introducing a singularity boundary condition 
\cite{Kunstatter:2012np} 
that mimics the Robin family of boundary conditions in
Schr\"odinger quantisation on the half-line~\cite{extensions,walton}: 
for the attractive Coulomb
potential and the attractive scale invariant potential, 
all these regularisations yield 
spectra that approximate the continuum quantum theory in the
relevant limit~\cite{coulomb,scale-invariant,Kunstatter:2012np}. 
Were this approximation to the continuum theory however 
found not to exist when there are 
lattice points arbitrarily close to the singularity 
but not at the singularity itself, 
one would be forced to conclude that
the regularisation mechanisms of 
\cite{Thiemann,coulomb,scale-invariant,Kunstatter:2012np} 
rely on a fine-tuning. Such 
fine-tuning would be undesirable 
in an implementation of fundamental 
discreteness in quantum theory. 

The purpose of this paper is to alleviate these fine-tuning concerns: 
we provide evidence that the singularity 
avoidance mechanisms based 
on a lattice point at the
classical singularity 
\cite{Thiemann,coulomb,scale-invariant,Kunstatter:2012np} 
\emph{are\/} in quantitative agreement with the 
spectrum that is obtained in the limit of lattice points
approaching the singularity. Our results can be understood as 
indirect support for the
Thiemann singularity avoidance prescription in loop quantum gravity 
\cite{Thiemann} 
and quantum field
theory~\cite{Hossain:2009vd,Hossain:2010wy,Hossain:2010eb,Husain:2013zda}.

Concretely, we 
consider the attractive Coulomb potential and the attractive scale
invariant potential on the discretised real line, first on an
equispaced lattice whose shift with respect to the origin approaches
zero, and then on a lattice that is equispaced except for two lattice
points that approach the origin symmetrically from both sides.  We
evaluate numerically the eigenenergies of the ground state and a
selection of low-lying excited states.  An analytic variational
argument shows that the ground state energy must decrease to negative
infinity, and for the non-equispaced lattice the same holds also for
the eigenenergy of the first excited state in the scale invariant
potential. However, the first key outcome is that all the higher
eigenvalues, within the range that our numerics is able to probe,
tend to finite limits, and in this limit the eigenvalues form
degenerate pairs.  The second key outcome is that all but the lowest
few of these pairs approach the eigenvalues of a discrete theory that
has a lattice point at the origin but whose singularity is regularised
by restriction to the odd parity
sector~\cite{scale-invariant}. Finally, the approach to degeneracy can
be reproduced from the discretised half-line quantum theory of
\cite{Kunstatter:2012np} by tuning the boundary condition parameter,
with a linear relation between the lattice shift and the boundary
condition parameter.

Within our potentials and lattices, there is hence quantitative 
agreement between the limit in which one or two lattice points 
approach the singularity and the  
regularised quantum theories in which a lattice point resides at the
singularity. A~qualitative discrepancy occurs only
with the one or two lowest-lying eigenenergies that descend to
negative infinity in the singular limit.

We begin in Section \ref{sec:polqm} by recalling relevant features of
quantisation on the discretised real line, for both of our lattices.
The spectra for the Coulomb potential and the scale invariant
potential are presented respectively in Sections \ref{sec:coulomb}
and~\ref{sec:scaleinv}, and these spectra are compared with the
discrete half-line theory in Section~\ref{sec:comparison}.  Section
\ref{sec:conclusions} gives a summary and concluding remarks.

We use dimensionless units, with $\hbar=1$. 
Complex conjugation is denoted by an overline.

\section{Quantum mechanics on the discretised real line} 
\label{sec:polqm}

In this section we outline our quantisation formalism 
on the discretised real line with our two lattices. 
For the equispaced lattice, 
a summary in the language of polymer quantum mechanics 
\cite{afw} is given in~\cite{Kunstatter:2012np}. 
We start by assuming that the potential is nonsingular 
and address the case of a singular potential 
in subsection~\ref{sec:singpot-latticetheory}.

\subsection{Real line: continuum} 

We consider a system whose classical phase space is 
$\BbbR^2 = \left\{ (x,p)\right\}$ with the Poisson bracket 
$\{ x,p\}=1$ and the Hamiltonian 
\begin{align}
H = p^2 + V(x) 
\ , 
\label{eq:class-Ham}
\end{align}
where the potential $V$ is real-valued and 
sufficiently well behaved. 

In standard Schr\"odinger quantisation, 
the classical Hamiltonian is promoted into the operator 
\begin{align}
{\widehat H}_{\text{S}} = -\partial_x^2  + \widehat V 
\ , 
\label{eq:Schr-Ham}
\end{align}
densely defined in the Hilbert space $L_2(\BbbR, dx)$, such that 
$\bigl(\partial_x^2 \psi\bigr)(x) = \psi''(x)$ and 
$\bigl(\widehat V\psi\bigr)(x) = V(x) \psi(x)$. 
If ${\widehat H}_{\text{S}}$ can be defined as a 
self-adjoint operator by a suitable choice of the domain, 
${\widehat H}_{\text{S}}$ generates unitarity evolution in 
$L_2(\BbbR, dx)$ by
Schr\"odinger's equation, 
$i\partial_t \psi = {\widehat H}_{\text{S}} \psi$. 

\subsection{Equispaced lattice} 
\label{subsec:equispacedtheory}

Our first lattice consists of the points $x_m = (m-\rho) \lscale$, 
where $m\in\BbbZ$ indexes the lattice points, the positive parameter 
$\lscale$ is the lattice spacing, 
and the real-valued parameter $\rho$ is such that the lattice point $x_0$ 
has been shifted left from the origin by $\rho\lscale$. 
We may assume $\rho \in [0, 1)$ without loss of generality. 

The Hilbert space is the space of two-sided sequences, 
$c := {(c_m)}_{m=-\infty}^\infty$, square summable in the inner product 
$(d, c) = 
\sum_m \overline{d_m} \, c_m$. 
We define the Hamiltonian 
${\widehat H}_{\text{eq}}$ by  
\begin{equation}
\bigl({\widehat H}_{\text{eq}} c\bigr)_m 
= \frac{2 c_m - c_{m+1} - c_{m-1}}{\lscale^2}
+ V(x_m) c_m 
\ , 
\label{eq:eqH-action}
\end{equation}
where the kinetic term is the standard three-point 
equispaced discretisation of~$-\partial_x^2$. 

${\widehat H}_{\text{eq}}$ is symmetric, and 
a study of the deficiency indices 
\cite{reed-simonII} shows that 
${\widehat H}_{\text{eq}}$ has at least one self-adjoint extension. 
If the set $\{ V(x_m) \mid m\in\BbbZ \}$ is bounded, the 
Kato-Rellich theorem 
\cite{reed-simonII}
can be applied as in 
\cite{scale-invariant} to show that ${\widehat H}_{\text{eq}}$ is essentially
self-adjoint. 

The special cases $\rho=0$ and $\rho=\frac12$ give lattices that are 
invariant under the reflection $x\mapsto-x$, 
the former with a lattice point at the origin, the latter 
with the origin half-way between two lattice points. 
When $V$ is even, we may choose $\rho \in [0, \frac12]$ 
without loss of generality, and the special cases 
$\rho=0$ and $\rho=\frac12$ then make 
${\widehat H}_{\text{eq}}$ parity invariant, 
so that the spectrum decomposes into the sector of even 
eigenfunctions and the sector of odd eigenfunctions. 
For $\rho \in (0, \frac12)$, ${\widehat H}_{\text{eq}}$ 
does not have a similar parity invariance for generic even~$V$, 
and the spectrum does not 
need to decompose into the even and odd sectors.

\subsection{Non-equispaced lattice} 
\label{subsec:nonequispacedtheory}

Our second lattice consists of the points 
\begin{align}
y_m = 
\begin{cases}
(m-1+\rho) \lscale, & \text{for $m >0$}, 
\\
(m-\rho) \lscale, & \text{for $m \le0$}, 
\end{cases}
\end{align}
where $m\in\BbbZ$ indexes the lattice points, the positive parameter
$\lscale$ is the spacing between adjacent lattice points except $y_0$
and~$y_1$, and the positive parameter $\rho$ is such that the spacing
between $y_0$ and $y_1$ is~$2\rho\lscale$.  Note that $y_m = x_m$ for
$m \le0$, and the lattice is invariant under the reflection $x \mapsto
-x$.

The Hilbert space is again the space of two-sided sequences, 
$c := {(c_m)}_{m=-\infty}^\infty$, now square summable in the inner product 
$(d,c) = 
\tfrac12 (1 + 2\rho) \bigl( \overline{d_0} \, c_0 
+ \overline{d_1} \, c_1 \bigr)
+ \sum_{m\ne0,1} \overline{d_m} \, c_m$, 
and the Hamiltonian ${\widehat H}_{\text{neq}}$ is defined by 
\begin{equation}
\bigl({\widehat H}_{\text{neq}} c\bigr)_m 
= 
\begin{cases}
\displaystyle \frac{2 c_m - c_{m+1} - c_{m-1}}{\lscale^2}
+ V(y_m) c_m
\ , 
& \text{for $m>1$ and $m<0$}, 
\\[2ex]
\displaystyle\frac{(1+2{\rho})c_1-2{\rho}c_2-c_0}{\rho(1+2{\rho})\bar{\mu}^2}
+V(y_1)c_1 
\ , 
& \text{for $m=1$},
\\[3ex]
\displaystyle\frac{(1+2{\rho})c_0-2{\rho}c_{-1}-c_1}{\rho(1+2{\rho})\bar{\mu}^2}
+V(y_0)c_0 
\ , 
& \text{for $m=0$}. 
\end{cases}
\label{eq:neqH-action}
\end{equation}
The kinetic term in \eqref{eq:neqH-action}
is the unique 
three-point discretisation of $-\partial_x^2$ 
that is exact for quadratic polynomials, 
and the weights of the $m=0$ and $m=1$ terms in the inner 
product have been chosen so that ${\widehat H}_{\text{neq}}$ is symmetric. 
It can be shown as above that ${\widehat H}_{\text{neq}}$ 
has at least one self-adjoint extension, 
and that ${\widehat H}_{\text{neq}}$ is essentially
self-adjoint if the set $\{ V(y_m) \mid m\in\BbbZ \}$ is bounded. 

As noted above, the lattice is invariant under the reflection 
$x\mapsto-x$. 
When $V$ is even, ${\widehat H}_{\text{neq}}$ is hence parity invariant, 
and the spectrum decomposes into the even and odd sectors. 
In the special case $\rho=\frac12$, the lattice coincides with the 
equispaced lattice with $\rho=\frac12$. 
In the limit $\rho\to0$, both $y_0$ and $y_1$ approach the origin.

\subsection{Singular potential on a lattice}
\label{sec:singpot-latticetheory}

We have assumed above that the potential $V$ has domain~$\BbbR$.  We
now turn to the case in which $V$ has a singularity at $x=0$ but is
defined elsewhere. The equispaced lattice theory of subsection
\ref{subsec:equispacedtheory} remains well defined provided
$\rho\in(0,1)$, and the non-equispaced lattice theory of subsection
\ref{subsec:nonequispacedtheory} remains well defined as it stands
since there $\rho$ is by construction positive.  Note that on both
lattices the distance from the origin to the closest lattice point(s)
equals~$\rho\lscale$.

Suppose hence that $\rho$ is positive. 
We consider the attractive Coulomb potential, 
$V(x) = -1/|x|$, and the attractive scale invariant potential, 
$V(x) = -\lambda/x^2$, 
where $\lambda$ is a positive constant. We may assume $0<\rho\le\frac12$. 
It follows as in 
\cite{scale-invariant} that ${\widehat H}_{\text{eq}}$
and ${\widehat H}_{\text{neq}}$ are essentially self-adjoint 
and bounded below, for the Coulomb potential by $-1/(\rho\lscale)$ 
and for the scale invariant potential by $-\lambda/(\rho^2\lscale^2)$. 

If the system has a ground state, the ground state energy 
can be bounded from above by a variational ansatz. 
For ${\widehat H}_{\text{eq}}$, the ansatz $c_m = \delta_{m,0}$ 
gives the upper bound $2/(\lscale^2)-1/(\rho\lscale)$ 
for the Coulomb potential and the upper bound 
$2/(\lscale^2)-\lambda/(\rho^2\lscale^2)$ 
for the scale invariant potential. 
For ${\widehat H}_{\text{neq}}$, 
the ansatz $c_m = \delta_{m,0} + \delta_{m,1}$ 
gives similarly the upper bounds 
$2/[(1+2\rho)\lscale^2]-1/(\rho\lscale)$ 
and 
$2/[(1+2\rho)\lscale^2]-\lambda/(\rho^2\lscale^2)$. 
It follows, for both ${\widehat H}_{\text{eq}}$ and~${\widehat H}_{\text{neq}}$, 
that the ground state energy decreases to negative infinity as $\rho\to0$. 

For ${\widehat H}_{\text{neq}}$, if the odd sector has
a lowest-energy state, we can obtain an upper bound for this
eigenenergy by the variational ansatz $c_m = \delta_{m,0} -
\delta_{m,1}$. For the scale invariant potential, the upper bound is
$2(1+\rho)/[\rho(1+2\rho)\lscale^2]-\lambda/(\rho^2\lscale^2)$,
showing that the lowest eigenenergy decreases to negative infinity as
$\rho\to0$. For the Coulomb potential, the upper bound is
$2(1+\rho)/[\rho(1+2\rho)\lscale^2]-1/(\rho\lscale)$. This shows that
the lowest eigenenergy decreases to negative infinity as $\rho\to0$
when $\lscale>2$, but does not guarantee such decrease when
$\lscale\le2$.

We shall see these phenomena in the numerical evaluation of the low
energy eigenvalues in Sections \ref{sec:coulomb} and
\ref{sec:scaleinv} below.

We note in passing that the singularity in both the attractive Coulomb
potential and the attractive scale invariant potential is so strong
that in the continuum quantum theory the positive half-line and the
negative half-line are decoupled from each
other~\cite{fewster,narnhofer}.  In a lattice quantum theory that
incorporates points from both halves of the real axis, this suggests
that a comparison to the continuum theory should only use
parity-invariant observables.  If this suggestion is adopted, the even
and odd sectors on a parity-invariant lattice become superselected, in
the sense that all observables map even states to even states and odd
states to odd states~\cite{Giulini:2007fn}.  The even and odd sectors
on our on-equispaced lattice can hence be viewed as superselection
sectors.  For a regular lattice with one point at the singularity,
this superselection terminology was adopted in~\cite{coulomb}.

\section{Coulomb potential} 
\label{sec:coulomb}

In this section we consider the attractive Coulomb potential. 
We first recall relevant facts about the continuum quantum theory 
and then analyse the discrete quantum theory on our two lattices. 

\subsection{Continuum}

We write the attractive Coulomb potential as $V(x) = -1/|x|$. This is
the theory of the spherically symmetric sector of the hydrogen atom in
Rydberg units, with $|x|$ being twice the Rydberg radial
coordinate~\cite{merzbacher}.

The singularity at $x=0$ is so strong that the 
positive and negative halves of the real axis decouple, 
and we may take the Hilbert space to be $L_2(\BbbR_+, dx)$. 
The self-adjoint extensions of 
${\widehat H}_{\text{S}}$ \eqref{eq:Schr-Ham} then form a 
$\text{U(1)}$ family, 
specified by a boundary condition at the origin~\cite{fewster}.  
The spectrum of each extension consists of the positive continuum and a 
countable set of negative eigenvalues. 
In the particular extension whose boundary condition 
is $\psi(0)=0$, the eigenvalues are $E = - 1/(4 s^2)$ with $s=1,2,3,\ldots\,$: 
this is the textbook quantisation 
of the spherically symmetric sector of the hydrogen
atom~\cite{merzbacher}. 

A~detailed technical analysis can be found in~\cite{fewster}. 

\subsection{Equispaced lattice}

We consider first the equispaced lattice. 
We look for the negative energy eigenvalues by solving the 
eigenvalue equation ${\widehat H}_{\text{eq}}\psi = E\psi$ numerically. 

The eigenvalue equation is a difference equation with a three-term
recurrence relation, and for negative $E$ the boundary condition is
that the solutions must decrease at both infinities sufficiently
rapidly to be normalisable.  Our numerical scheme is adapted from that
in~\cite{coulomb}.  Given an $E<0$ and a (large) positive integer
cut-off~$m_0$, we first use the method of \cite{coulomb} to find a
solution $\bigl\{c^{(+)}_m \mid m=0,1,\ldots,m_0\bigr\}$ that is
exponentially suppressed at large positive~$m$, and we then similarly
find a solution $\bigl\{c^{(-)}_m \mid
m=-m_0,-m_0+1,\ldots,0,1\bigr\}$ that is exponentially suppressed at
large negative~$m$.  The eigenvalues are those $E$ for which these two
solutions coincide at $c_0$ and $c_1$ up to normalisation. The
condition that determines the eigenvalues is hence $c^{(-)}_0
c^{(+)}_1 - c^{(-)}_1 c^{(+)}_0 =0$, which we solve by the shooting
method.  Numerical accuracy is monitored by increasing $m_0$ until the
results stabilise.

\begin{figure}[p]
\begin{center}
\includegraphics[width=0.49\textwidth]{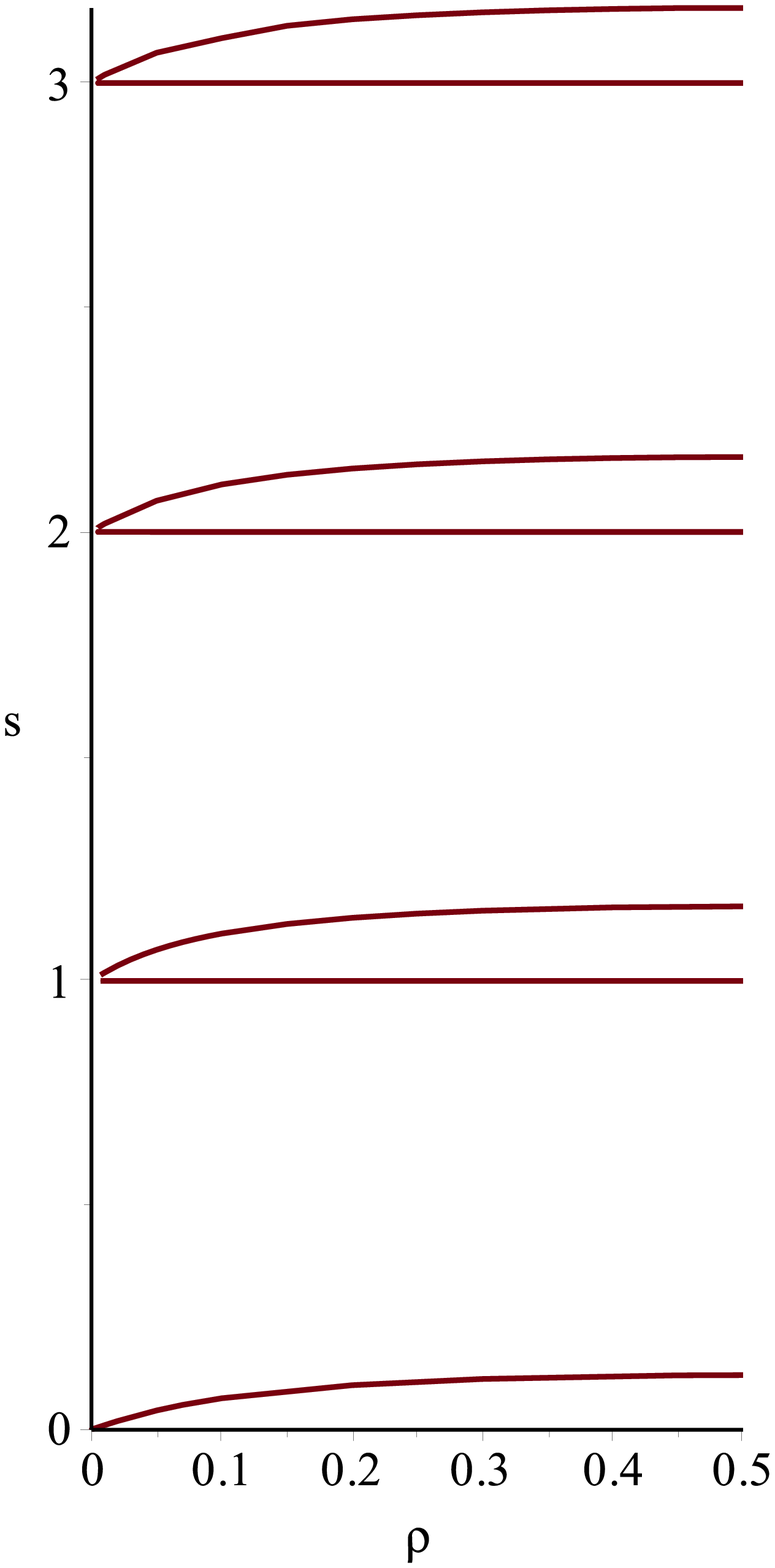}%
\includegraphics[width=0.49\textwidth]{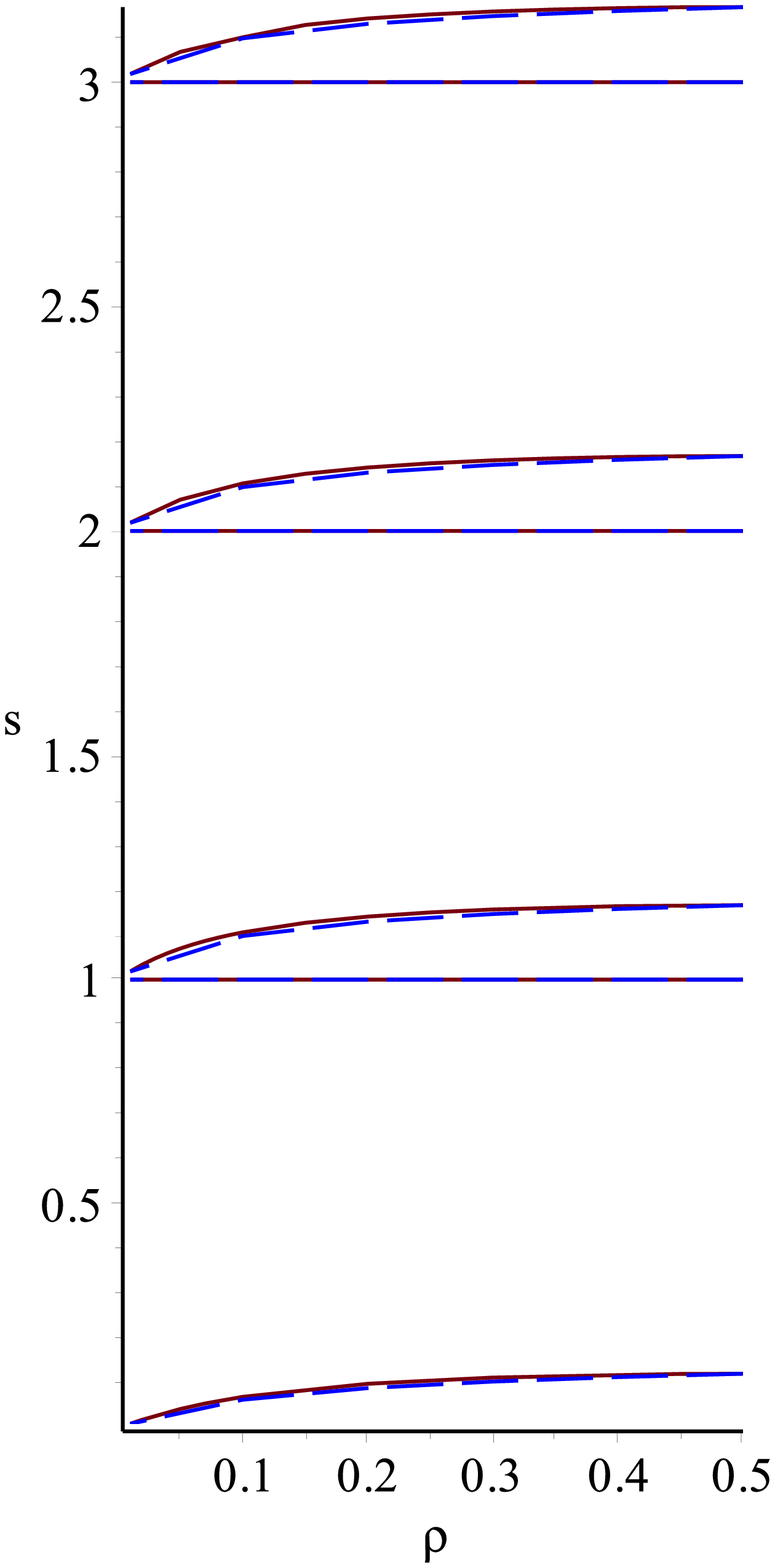}
\caption{Eigenvalues for the Coulomb potential with $\lscale = 0.01$. 
The quantity plotted is $s = 1/\sqrt{-4E}$. 
The equispaced lattice is on the left, 
showing the lowest seven eigenvalues as a function of~$\rho$. 
The plot on the right reproduces the data from the left as solid 
(red) lines and superposes the 
corresponding data for the non-equispaced 
lattice as dashed (blue) lines.%
\label{fig:coulomb-s}}
\end{center}
\end{figure}

Motivated by the eigenvalues of the continuum theory, we parametrise
the lattice eigenvalues as $E = - 1/(4 s^2)$ where $s>0$.  Figure
\ref{fig:coulomb-s} shows the lowest seven eigenvalues as a function
of $\rho$ for $0.004\le \rho \le \frac12$ when $\lscale=0.01$. Probing
arbitrarily small values of $\rho$ is not numerically possible, but
the plot presents strong evidence for the asymptotic behaviour as
$\rho\to0$.  The eigenvalues split into two alternating sets, which we
call the A set and the B set, such that the A set includes the
ground state.  In the A set, the eigenvalues approach their asymptotic
limits from above, and for the ground state this limit appears to be
at negative infinity, in agreement with the bound given in
subsection~\ref{sec:singpot-latticetheory}.  In the B set, the
eigenvalues approach their asymptotic limits from below. In the limit,
the excited state eigenvalues appear to form degenerate pairs, to the
numerical accuracy that we have been able to probe, stabilising at
values that are slightly above those of the continuum theory with the
textbook boundary condition.

For $\lscale=0.1$, the behaviour of the
eigenvalues is qualitatively similar, and the 
$\rho\to0$ limits of the excited 
eigenvalues are slightly further 
above the continuum textbook eigenvalues. 
This could have been expected on the grounds that if we choose 
$\rho=0$ at the start and regularise the singularity at 
the origin by restriction to the odd parity sector, 
the $\lscale\to0$ limit 
appears to converge to the continuum 
textbook theory~\cite{scale-invariant}. 
At $\lscale<0.01$ the numerics becomes significantly slower 
and we have not examined this regime.

\subsection{Non-equispaced lattice}

On the non-equispaced lattice, the Hamiltonian ${\widehat
  H}_{\text{neq}}$ is parity invariant, and the spectrum hence
decomposes into the even and odd sectors for all values
of~$\rho$. In the numerical solution of the eigenvalue equation
${\widehat H}_{\text{neq}}\psi = E\psi$, we may therefore proceed as
with the equispaced lattice to compute first the solution
$\bigl\{c^{(+)}_m \mid m=0,1,\ldots,m_0\bigr\}$ for fixed $E$
and~$m_0$, and then find the eigenvalues as those $E$ for which $c_0 =
\pm c_1$, where the upper (lower) sign gives the even (odd)
sector.

A~plot of the lowest seven eigenvalues with $\lscale=0.01$ 
is shown in Figure~\ref{fig:coulomb-s}. 
For $\rho=\frac12$, the equispaced and non-equispaced lattices
coincide, and we find that the $A$ and $B$ eigenvalue sets found above
coincide respectively with the even and odd sectors.
As $\rho$ decreases, the coincidence is no longer precise, but it
continues to hold to a good approximation: the even 
sector eigenvalues lie closely below the set A eigenvalues, while the
odd sector eigenvalues lie closely above the set B
eigenvalues. 
In particular, the ground state eigenvalue 
again descends to the negative infinity as $\rho\to0$, 
in agreement with the bound given 
in subsection~\ref{sec:singpot-latticetheory}, 
and the excited state eigenvalues are sandwiched 
between the equispaced lattice set A and set B eigenvalues, 
forming degenerate pairs as $\rho\to0$.

\section{Scale invariant potential} 
\label{sec:scaleinv}

In this section we consider the attractive scale invariant potential. 
We again first recall relevant facts about the continuum quantum theory 
and then analyse the discrete quantum theory on our two lattices. 

\subsection{Continuum}

We write the attractive scale invariant potential as $V(x) =
-\lambda/x^2$, where $\lambda$ is a positive constant.  The positive
and negative halves of the real axis again decouple, and we may take
the Hilbert space to be $L_2(\BbbR_+, dx)$.  The self-adjoint
extensions of ${\widehat H}_{\text{S}}$ \eqref{eq:Schr-Ham} are
classified in~\cite{narnhofer}: early analyses were given in
\cite{case,frank-potential} and a review with further references can
be found in~\cite{scale-invariant}.  In the regime $\lambda>\frac14$,
which we shall consider in the lattice theories below, the spectrum
consists of the positive continuum and a countable set of negative
eigenvalues, given by
\begin{align}
E_n = E_0 \exp\bigl(-2\pi n /\sqrt{\lambda-(1/4)}\,\bigr)
\ , 
\label{eq:scaleinv-cont-evals}
\end{align}
where $n\in\BbbZ$ and the negative constant $E_0$ is determined by the
boundary condition.

\subsection{Equispaced lattice}

On the equispaced lattice we solve the eigenvalue equation 
${\widehat H}_{\text{eq}}\psi = E\psi$ 
numerically by the same method as for
the Coulomb potential.  Because the discretisation preserves the scale
invariance of the continuum theory, the lattice spacing $\lscale$
enters the eigenvalues only as the overall multiplicative factor
$1/\lscale^2$. Motivated by the continuum theory
eigenvalues~\eqref{eq:scaleinv-cont-evals}, we parametrise the lattice
eigenvalues by 
$\lscale^2 E= - \exp(-s)$ where $s\in\BbbR$.

\begin{figure}[p]
\begin{center}
\includegraphics[width=0.49\textwidth]{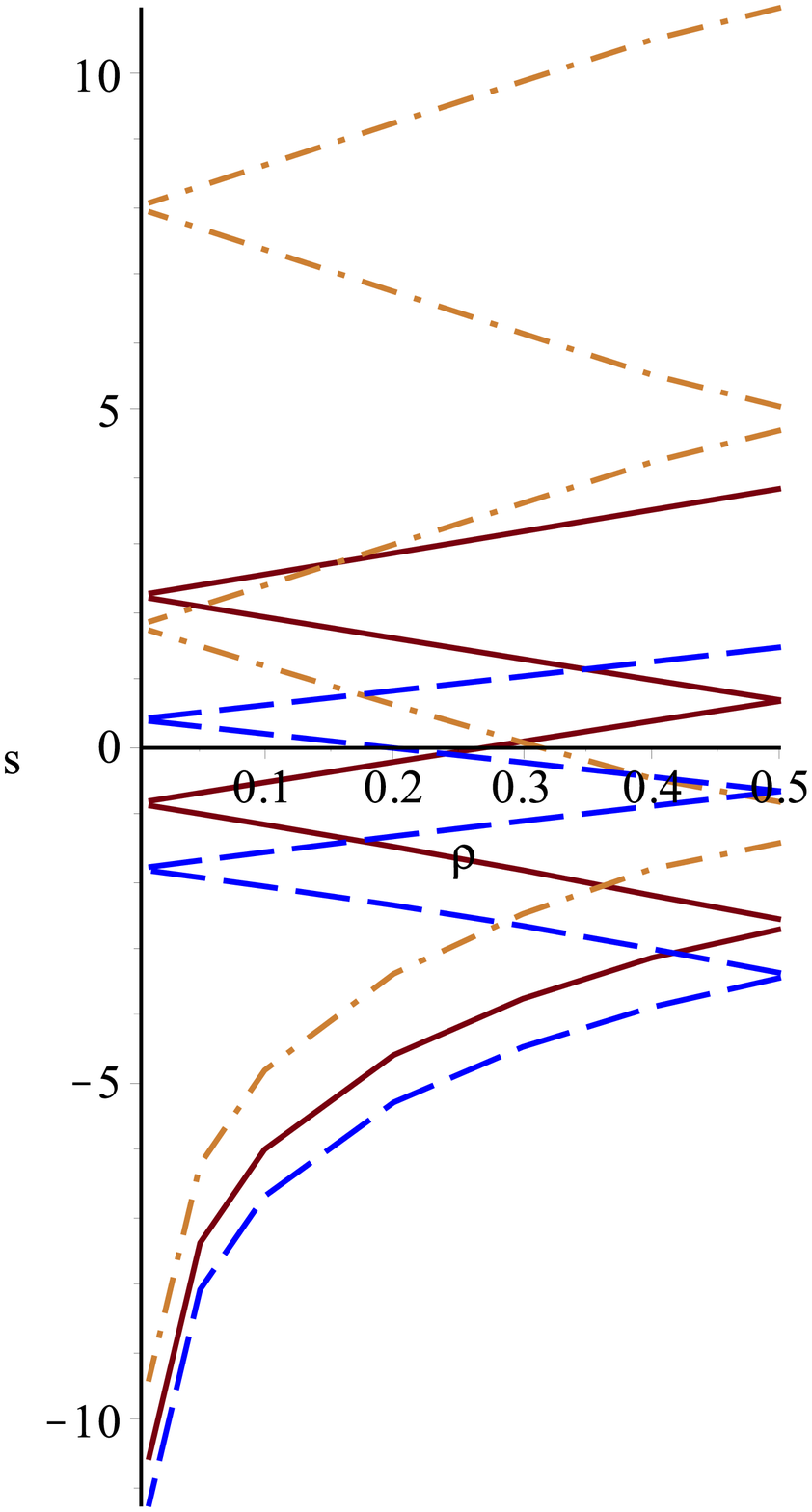}%
\includegraphics[width=0.49\textwidth]{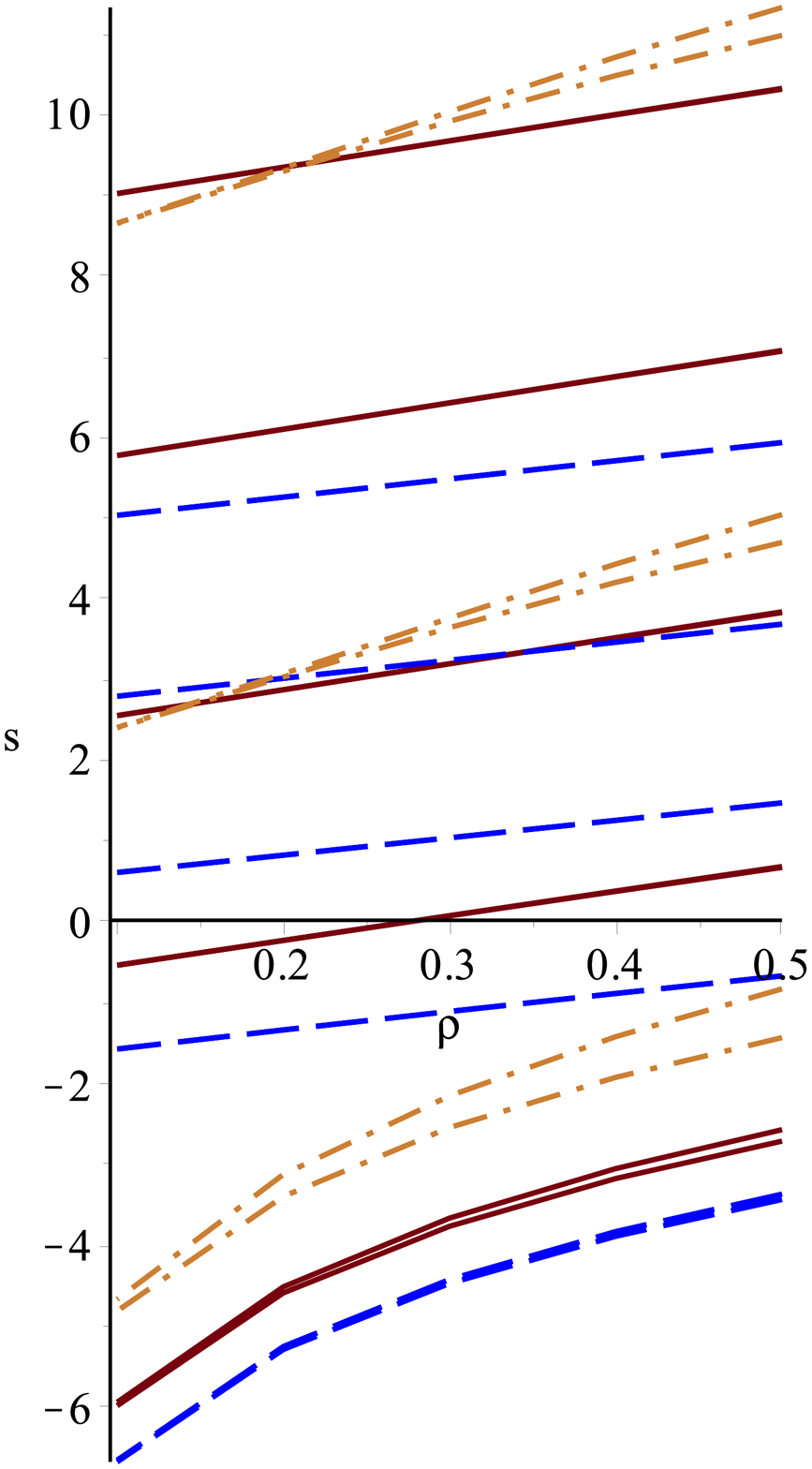}
\caption{Eigenvalues for the scale invariant potential with 
$\lambda=1.25$ (dot-dashed, gold), 
$\lambda=4$ (solid, red) 
and 
$\lambda=8$ (dashed, blue). The quantity plotted is $s = - \ln(-\lscale^2 E)$. 
The equispaced lattice is on the left, showing 
the lowest five eigenvalues for each $\lambda$ as a 
function of~$\rho$. 
The non-equispaced lattice is on the right, showing 
the lowest six eigenvalues for $\lambda=1.25$ 
and the lowest ten eigenvalues for $\lambda=4$ 
and 
$\lambda=8$. In the plot on the right, 
the resolution does not suffice
to show the descent of the lowest-lying pair towards negative infinity as
$\rho\to0$, and the resolution separates for $\lambda=4$ 
only the lowest-lying pair of eigenvalues 
and for $\lambda=8$ none of the pairs of eigenvalues.%
\label{fig:scaleinv-s}}
\end{center}
\end{figure}

Figure \ref{fig:scaleinv-s} shows the lowest five
eigenvalues as a function of $\rho$ for $0.01\le \rho \le \frac12$,
with $\lambda=1.25$, $\lambda=4$ and $\lambda=8$.  
In the $\rho\to0$ limit, the behaviour is qualitatively similar to
that with the Coulomb potential. The eigenvalues split into
alternating A and B sets, tending to their limits respectively from
above and from below. The only eigenvalue that does not tend to a
finite limit is the ground state, which descends to negative infinity, 
in agreement with the bound 
given in subsection~\ref{sec:singpot-latticetheory}, 
and the higher eigenvalues form pairs that become degenerate,  
to the accuracy that we have been able to probe.

\subsection{Non-equispaced lattice}

On the non-equispaced lattice, the eigenvalues break into the even and
odd sectors, and they can be investigated by the same numerical
methods as with the Coulomb potential.

As $\rho\to0$, the lowest eigenvalue in each of the two sectors
descends towards negative infinity, in agreement with the bounds given
in subsection~\ref{sec:singpot-latticetheory}.  Our numerical evidence
is inconclusive as to how close to each other the two eigenvalues will
remain in the final stages of this descent.

For $\rho\le 10^{-8}$, the higher eigenvalues in the two sectors
coincide to at least 14 decimal places in their $s$-values, and they
further coincide with the corresponding set A eigenvalues of the
equispaced lattice within our numerical accuracy.  In particular, all
these eigenvalues tend to finite limits as $\rho\to0$.

A plot of the low eigenvalues for $\lambda=1.25$, $\lambda=4$ and
$\lambda=8$ is shown in Figure~\ref{fig:scaleinv-s}.

\section{Comparison with discrete half-line} 
\label{sec:comparison}

In all the cases analysed above, we have seen that in the limit
$\rho\to0$ the spectrum breaks into two subsets, each of which has a
strong qualitative resemblance with the eigenvalues obtained by
regularising the singularity of the potential, whether by explicitly
modifying the functional form of the
potential~\cite{coulomb,scale-invariant}, by a parity
argument~\cite{scale-invariant}, or by formulating the lattice theory
on a half-line with a Robin-type boundary condition at the
singularity~\cite{Kunstatter:2012np}.  The only exceptions in this
qualitative resemblance occur in the one or two states of lowest
energy.

We shall now investigate the resemblance quantitatively. 

Among the regularised theories, we consider the theory defined on the
discrete half-line~\cite{Kunstatter:2012np}.  The lattice is
semi-infinite, the lattice points are at $z_m = m\lscale$ with
$m=1,2,3,\ldots$, and the inner product reads $(d, c) =
\sum_{m=1}^\infty \overline{d_m} \, c_m$. The Hamiltonian
$\widehat{H}_\alpha$ is given by
\begin{align}
\bigl(\widehat{H}_\alpha c\bigr)_m 
= 
\begin{cases}
{\displaystyle \frac{2 c_m - c_{m+1} - c_{m-1}}{\lscale^{2}} + V(m\lscale) c_m}
\ , 
& 
\text{for $m>1$,} 
\\[2ex]
{\displaystyle \frac{(2-\alpha) c_1 - c_{2}  }{\lscale^{2}} + V(\lscale) c_1}
\ , 
& 
\text{for $m=1$.} 
\end{cases} 
\label{eq:fHalpha-action}
\end{align}
where $\alpha$ is a real-valued parameter.  Note that as $z_0=0$ is
not part of the lattice, the value of the potential at the singularity
does not enter the Hamiltonian. 

The parameter $\alpha$ may be thought of as specifying a Robin-like
boundary condition at the fictitious lattice point $m=0$, with the
special case $\alpha=0$ corresponding to the Dirichlet-like condition
$c_0=0$. The theory with $\alpha=0$ is equivalent to a theory
that is defined on an equispaced lattice over the full real line, with
one lattice point at the singularity, but regularised by restriction
to the odd parity sector~\cite{scale-invariant}. This observation will
be significant below.


We pose the following questions. For the equispaced lattice, can the
set A (respectively set B) eigenvalues be reproduced from the discrete
half-line theory with some choice of $\alpha$ as a function of~$\rho$
in the limit $\rho\to0$? For the non-equispaced lattice, can the
eigenvalues in the even (respectively odd) sector be
reproduced from the discrete half-line theory with some choice of
$\alpha$ as a function of~$\rho$ in the limit $\rho\to0$?

We consider the Coulomb potential and the scale invariant potential in turn.

\subsection{Coulomb potential}
\label{subsec:alpha-coulomb}

For the Coulomb potential, we have carried out a numerical analysis 
on the equispaced lattice with $\lscale=0.01$ and $\lscale=0.1$. 

We find that the answer is affirmative: the eigenvalues in set A and
set B can be reproduced from the discrete half-line theory in the
limit $\rho\to0$, and for each of the sets the dependence of $\alpha$
on $\rho$ fits the linear relation $\alpha = K\rho$ to a good
approximation when $\rho \leq 10^{-5}$. The values of the coefficient
$K$ for each set and each value of $\lscale$ are shown in
Table~\ref{tab:coulomb-rhovsalpha-eq}. The only eigenvalue that does
not fit the pattern is the ground state.

Note that $\alpha \to 0$ as $\rho\to0$. This means that in the
$\rho\to0$ limit, the eigenvalue pairs approach the eigenvalues in the
equispaced lattice theory in which one lattice point is at the
singularity but the theory is regularised by restriction to the odd
parity sector~\cite{scale-invariant}.

We have not carried out a similar computation on the non-equispaced
lattice. However, the sandwiching of the nonequispaced lattice
eigenvalues between the equispaced lattice eigenvalues, shown in
Figure~\ref{fig:coulomb-s}, strongly suggests that a similar
correspondence holds, so that the even sector and the
odd sector can be matched to the discrete half-line
theory in the $\rho\to0$ limit, each with a linear relation between $\rho$
and~$\alpha$. The sandwiching
gives bounds on the coefficients in these linear
relations, in terms of the coefficients shown in
Table~\ref{tab:coulomb-rhovsalpha-eq}. In particular, the sandwiching
shows that in the $\rho\to0$ limit the eigenvalue pairs again approach
the eigenvalues in the equispaced lattice theory with a lattice point
at the singularity but regularised by restriction to the odd parity
sector.

\begin{table}[t]
\footnotesize
\centering
\begin{tabular}{l|r r}
\hline\hline
                        & A $\phantom{200.0}$     & B $\phantom{200}$ \\
\hline
$\lscale = 0.1 \vphantom{A^{A^{A}}}$ & $-20.05514$  & $0.05513$ \\
$\lscale = 0.01$                     & $-200.00527$ & $0.00505$ \\[.5ex]
\hline\hline
\end{tabular}
\caption{Coulomb potential on the equispaced lattice. 
The table shows the coefficient $K$ in the asymptotic small $\rho$ 
relation $\alpha = K \rho$ for the set A and set B eigenvalues, 
with $\lscale = 0.1$ and with $\lscale = 0.01$.%
\label{tab:coulomb-rhovsalpha-eq}}
\end{table}

\begin{table}[t]
\footnotesize
\centering
\begin{tabular}{l|r r}
\hline\hline
                        & A $\phantom{200.000}$     & B $\phantom{20000}$ \\
\hline
$\lambda = 4 \vphantom{A^{A^{A}}}$ & $-17.44175$  & $17.44175$ \\
$\lambda=8$                        & $-125.00085$ & $125.00085$ \\[.5ex]
\hline\hline
\end{tabular}
\caption{Scale invariant potential on the equispaced lattice. 
The table shows the coefficient $K$ in the asymptotic small $\rho$ 
relation $\alpha = K \rho$ with $\lambda=4$ and $\lambda=8$, for 
set A and set~B\null.%
\label{tab:scaleinv-rhovsalpha-eq}}
\end{table}

\subsection{Scale invariant potential}
\label{subsec:alpha-scaleinv}

Consider the scale invariant potential with the equispaced lattice. 
For $\rho \leq 10^{-4}$, we find that
both the set A eigenvalues and the set B eigenvalues can be reproduced
from the discrete half-line theory with the linear relation
$\alpha=K\rho$, provided we exclude the three lowest eigenvalues
from the A set and the two lowest eigenvalues from the B set. The
values of the coefficient $K$ are shown in Table
\ref{tab:scaleinv-rhovsalpha-eq} for each set, with $\lambda=4$ and
$\lambda=8$. The coefficients for the two sets 
differ only in the overall sign, within our numerical
accuracy: we have not investigated analytically whether this
difference by only the sign might be exact.

On the non-equispaced lattice, we saw above that in each
sector all eigenvalues except the lowest one coincide
with those of the equispace lattice A set as $\rho\to0$, to high
precision.  We hence have again a correspondence to the half-line
theory, with the linear relation $\alpha=K\rho$, and the values of the
coefficient $K$ are given by column A in
Table~\ref{tab:scaleinv-rhovsalpha-eq}.

Note again that $\alpha\to0$ as $\rho\to0$. Hence, with the exception
of the lowest few eigenvalues, the eigenvalue pairs approach the
eigenvalues in the equispaced lattice theory with a lattice point at
the singularity but regularised by restriction to the odd parity
sector~\cite{scale-invariant}.

\section{Conclusions} 
\label{sec:conclusions}

In this paper we have investigated nonrelativistic quantum mechanics
on the discretised real line in two classically singular potentials,
the attactive Coulomb potential and the attractive scale invariant
potential. We considered an equispaced lattice and a lattice that is
equispaced except for one shorter interval that ensconces the
classical singularity, and we analysed the spectrum in the limit in
which one or two lattice points approach the singularity, by numerical
evaluation of the bound state eigenvalues. We found that while one or
two of the lowest energy eigenstates descend to negative infinity in
this limit, the remaining eigenvalues tend to finite limits that form
degenerate pairs and are close to the eigenvalues of the continuum
theory, and also close to the eigenvalues of discrete theories in
which the singularity has been regularised by the Thiemann mechanism
introduced in loop quantum gravity~\cite{Thiemann}, by a parity
argument~\cite{scale-invariant}, or by a discrete version of the
continuum Robin boundary condition~\cite{Kunstatter:2012np}. We in
particular established that the approach to degeneracy can be
quantitatively reproduced from the discrete half-line theory by tuning
the boundary condition parameter therein to be a linear function of
the distance from the singularity to the closest lattice point(s).

These results bear witness to unanimity amongst the mechanisms by
which a classically singular continuum theory becomes nonsingular on
quantisation, whether the quantisation is built on a continuous
configuration space or on a discrete configuration space. If the
discreteness is thought of as a pragmatic tool, as an approximation to
a `true' quantisation with a continuous configuration space, the
results are evidence that the various discrete treatments of the
classical singularity yield compatible results and efficient
approximations. If the discreteness is thought of as fundamental, the
results are evidence that the core properties of the discrete theory
do not rely on fine-tuning in the way in which the discreteness
implements singularity avoidance. From this viewpoint, our results
provide indirect support for the Thiemann singularity avoidance
prescription in loop quantum gravity \cite{Thiemann} and quantum field
theory~\cite{Hossain:2009vd,Hossain:2010wy,Hossain:2010eb,Husain:2013zda}.

For the Coulomb potential, the cases of one or two lattice points
approaching the singularity were qualitatively very similar, even in
the lowest eigenvalues.  For the scale invariant potential, by
contrast, approaching the singularity with two lattice points affected
the lowest eigenvalues significantly more strongly than with just one
lattice point.  This could perhaps have been expected on the grounds
that the singularity in the scale invariant potential is stronger than
in the Coulomb potential.  There would be scope for a systematic study
of this phenomenon within a wider range of singular potentials,
power-law and beyond, and within a wider variety of lattices.

Finally, in this paper we have considered only the bound state part of
the spectrum. It would be equally interesting to study the behaviour
of travelling wave packets \cite{throatb} in the limit where one or
more lattice points approach the singularity: does the unbounded
descent of the lowest one or two eigenvalues leave a footprint in the
reflection and transmission of waves? If yes, can the footprint be
reproduced from a manifestly nonsingular theory, or could it possibly
contain evidence of pathology, detectable by low energy observations?

\section*{Acknowledgements}

We thank Chris Fewster for insightful comments, including the
suggestion to use a variational estimate for the ground state
energy.
We also thank 
Viqar Husain and Gabor Kunstatter for 
helpful discussions and correspondence. 
S.P.P. and M.D.W. were supported by EPSRC undergraduate research
bursaries. 
J.L. was supported in part by 
STFC 
(Theory Consolidated Grant ST/J000388/1).

\end{document}